weeteq

*Communication*

# Intelligent Real-Time Transient Response Optimization

Anahita Kanade[1], Djan Tanova[1], Steve Conner[1] and Taner Dosluoglu[1],* *Member, IEEE*

***Abstract*—Use of artificial intelligence in motor control applications that can be deployed within the microcontrollers need to comply with real-time demands of motor control systems. A supplementary controller approach that can be integrated within existing microcontrollers is presented. The proposed approach is implemented on a weeteq motor control integrated circuit and tested in the lab. A complete unsupervised motor control deployment solution was developed, and the real-time system response correction demonstrated under dynamic loads measured with and without the supplementary controller with embedded artificial intelligence. The solution provides 100% coverage of real-time data at control loop sample rate and model inference period between 100usec and 1msec. The importance of latency for reduction of dynamic regulation margin during transient response is demonstrated with up to 68% reduction of the system response error. A novel key performance indicator based on principal components transform is introduced that provides a quantitative figure of merit for improvement of the transient response, which includes the dynamic regulation margin, stability considerations and iterative improvements of consecutive regression model outputs. The significant events related to dynamic changes in the equipment and external operating conditions are detected at milliseconds resolution and recorded as highly compressed vectors representing deviation of the system response from linear steady state conditions. The resolution in time and accuracy of this vector data will enable a new level of system level optimization that has not been possible using the time series data from IoT sensors in current equipment health monitoring solutions.**

## I. INTRODUCTION

The application of artificial intelligence to motor control has been a topic of research for four decades. While the initial work has focused on replacing the linear control systems with machine learning models, alternative methodologies, such as supplementary control offers better value that can currently be implemented using existing microcontrollers with competitive design, integration and deployment costs. A key insight from a review of recent studies on machine learning and nonlinear dynamics can be summarized as how the nonlinearity emerges as a structural property of the underlying dynamics when linear constraints are removed rather than an increased model complexity [1]. Incorporation of state-space modelling, multi-physics frameworks into simulation platforms has helped optimization of motor design and thermal control but the challenge remains for addressing nonlinearities of magnetic materials within the constraints of real-time simulations in IoT enabled systems [2]. The practical implementation of artificial intelligence algorithms in the field also involves complexity, primarily due to the real-time demands of motor control systems. AI models, especially deep learning models, may require substantial processing time, making it crucial to optimize models to meet real-time requirements [3].

In this paper, we focus on a solution that can already be integrated within existing microcontrollers. The approach has been previously published [4] and we present the results from a motor controller Integrated Circuit (IC) we have released earlier this year as a demonstration and development platform [5]. Motor controller IC consists of an FOC controller (main controller) and a supplementary controller that generates vectors based on deviation of the system response from an ideal response. In essence, the supplementary control block captures non-linearities and gain/bandwidth constraints of the main controller (designed for optimal control) and identifies changes in the operating conditions beyond the response of the controller. It allows simplification in the design where complex state transitions can be addressed within the supplementary block. In addition to the motor controller IC, a test rig was developed (Figure 1) to provide precise and repeatable torque control using a second motor (generator) mechanically connected to the motor under test providing dynamic adjustment of torque. Dynamic torque transients with less than 5% deviation from benchmarks was demonstrated for WLTP (Worldwide Harmonized Light Vehicles Test Procedure) compliant testing and demonstration using IIR notch filters to mitigate resonance and ensuring robust transient response accuracy for real-world validation and performance enhancement under customer specified use cases.

## II. MATERIALS AND METHODS

Supplementary controller refers to Ultra Edge® technology which is a combination of a methodology for developing embedded solutions within a microcontroller architecture and corresponding implementation on a reference design. Use of supplementary controller is critical in reducing the latency of the model inference for detecting significant events in real-time.

This paper was originally submitted on 7 August 2026 for review. The research was partially funded by Scottish Enterprise, Smart Scotland Grant number SMART/2324/018 23/9091.

All authors are with weeteq LTD, Glasgow, UK *(Corresponding author: Taner Dosluoglu*, taner@weeteq.com.

weeteq

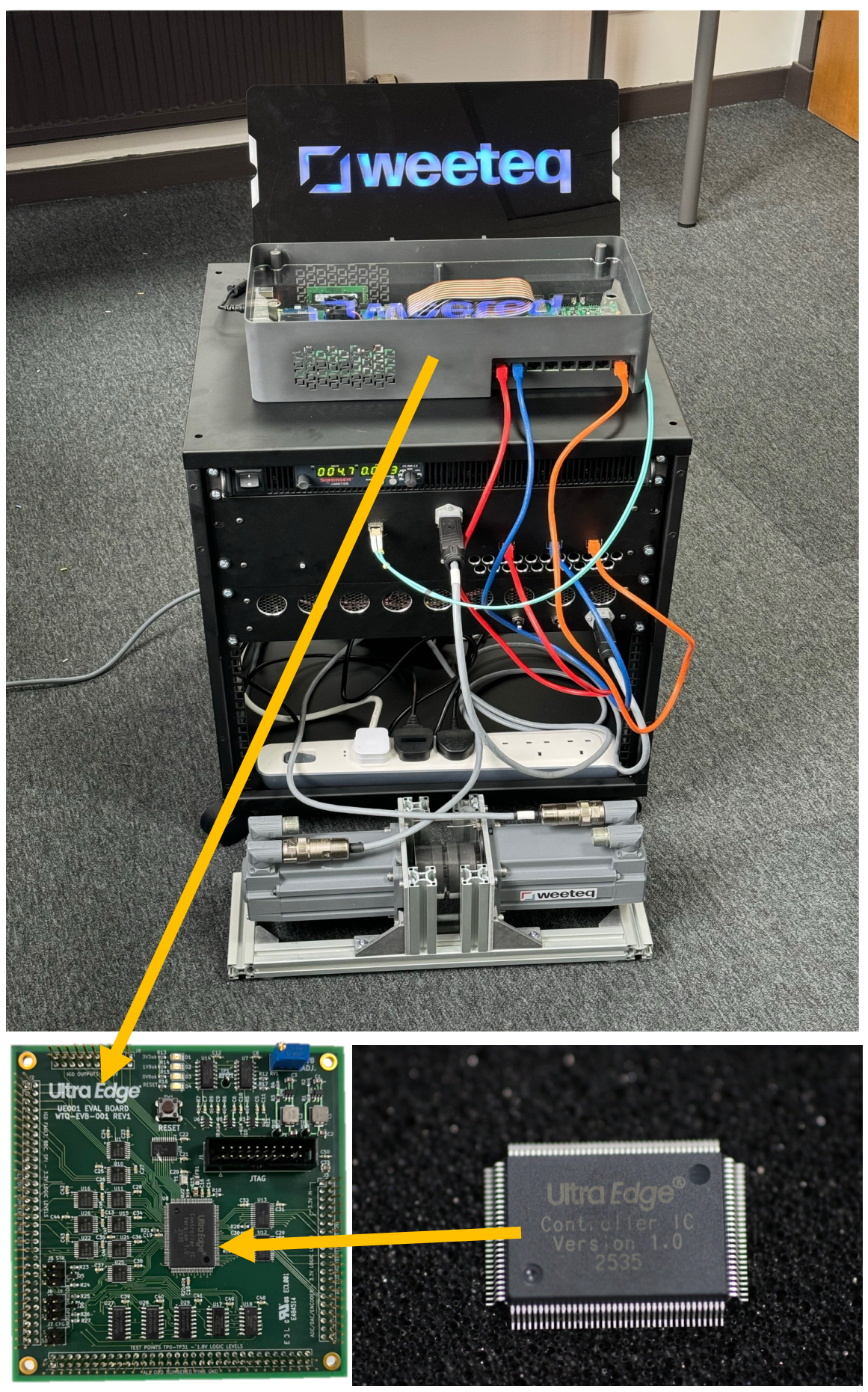

**Fig. 1.** Test-rig with coupled motors on the ground. The FPGA based box at the top is replaced by motor controller IC evaluation board shown in bottom left image.

We have limited our discussions to the methodology and functional overview of the solution. The results will be presented based on a reference design that we have implemented on a FPGA platform as well as our own silicon integrated circuit based on ARM Corstone-300 platform that includes ARM Cortex-M55 CPU core and ARM Ethos-U55 NPU core.

The data flow diagram is provided in Figure 2 and defines integration of Ultra Edge® supplementary controller within closed loop control with real-time inputs and outputs sampled at control loop data sample rates. The term control loop data sample rate is used to define processing 100% of the digital data available to the main controller of the application and will be referring to ADC sample rate. The inference model time period depends on the processing power of CPU/NPU and the model size used in the implementation. The data presented in this paper was obtained with model inference time period in the range of 100usec to 1msec.

Pre-processing and post-processing blocks handle the data rate conversion between the inference model time period and control loop data sample rate, normalization of data for corresponding digital formats, and apply various digital filter functions as needed for pre/post-processing.

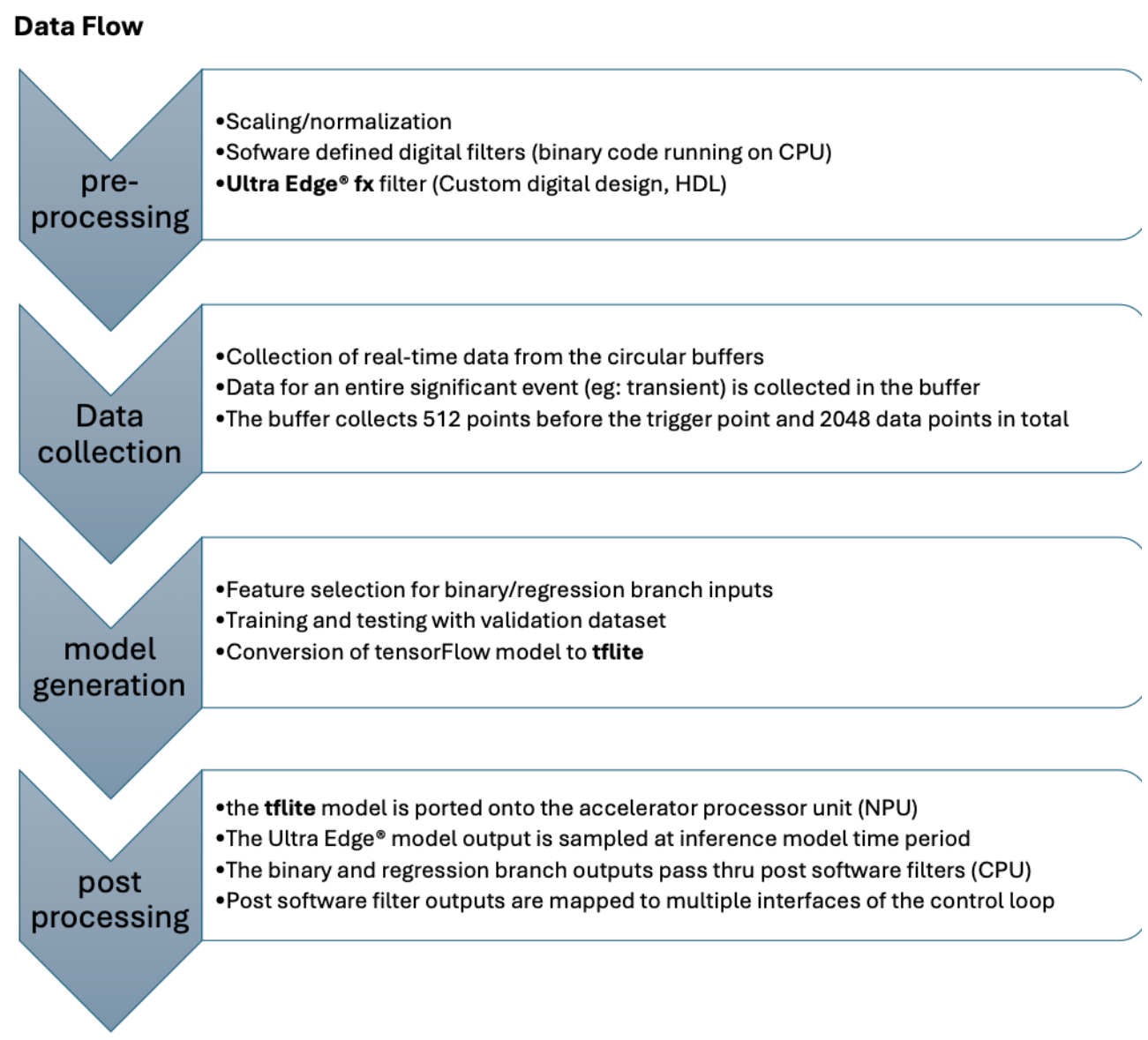

**Figure 2.** Data Flow for model generation

For the pre-processing block digital filters are used for measurement noise suppression and for improving the correlation between model inputs and significant events ("Ultra Edge fx" filter). Post-processing block digital filters provide extrapolation of inference model output to outer loop and inner loop inputs of the main controller. For real-time operation, model inferencing replaces the data collection and model generation steps of Figure 2 and the inference is applied to complete control loop data available for each inference model time period on a rolling basis. For a nominal 200usec inference model time period this is 4 samples at 20kHz ADC sample rate.

The model generation methodology is designed to allow unsupervised operation for the entire flow. This is shown in Figure 3 where the baseline model is customized at start-up and re-calibrated during the operation as needed. This requires definition of unsupervised significant event detection and performance analysis with available on-chip resources. The architecture is based on circular buffers that store required data in real-time, with control logic that freezes the circular buffers when a significant event is detected and performs subsequent analysis using the stored data.

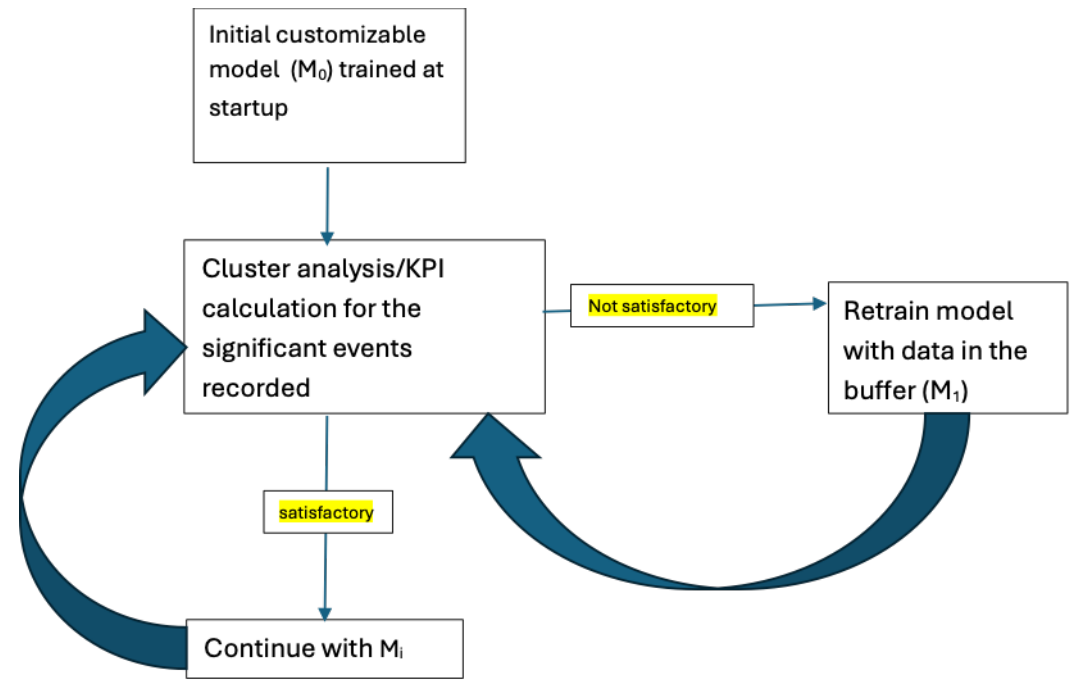

**Figure 3.** On-chip model generation.

weeteq

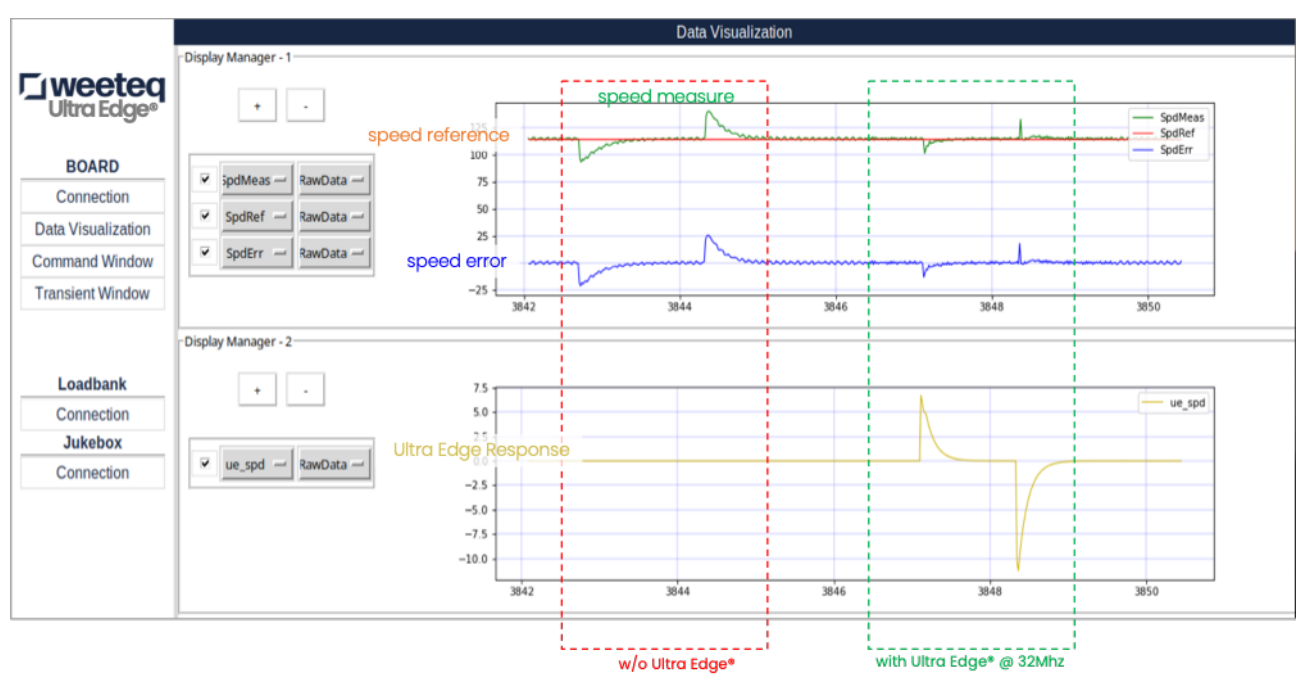

**Figure 4.** FPGA platform running at 32MHz comparison with and without Ultra Edge.

The normalization of the model inputs and gain/bandwidth constants are obtained during the startup operation (in parallel with the autotune function of the main controller) for the baseline model and pre/post-processing blocks. We have currently limited the on-chip training algorithm during the operation; to the last two layers of the model due to current resource constraints of the microcontrollers used.

The model consists of input layer, time2vec layers, output layers, LSTM layer, and hidden layers. Dropout Layers are also added to prevent overfitting. Binary classification and regression functions are performed simultaneously as two parallel branches with separate loss functions. Binary branch is used for significant event detection, and regression branch provides statistically valid prediction intervals and quantifies the un-certainty of the predictions with iterative convergence. The following libraries are used for model generation: Keras, Scikit-learn, TensorFlow, Pandas, Numpy, Tflite2onnx.

Torque changes are controlled independently by a dynamic torque control module developed in-house using a Raspberry Pi 3B which is running Volumio, a lightweight music player operating system. This system generates PWM signals that control the time duration of power devices connected to a resistor load bank for a motor used as a generator. Therefore providing torque by adjusting the resistance of the electrical load. This system is referred to as “weeteq jukebox” as it allows speed and torque scenarios to be “played” by the dynamic torque control module. In the lab setup, two 480V Allen Bradley MPL-B330P-MJ72AA motors were used, one acting as the driver and the other acting as the load.

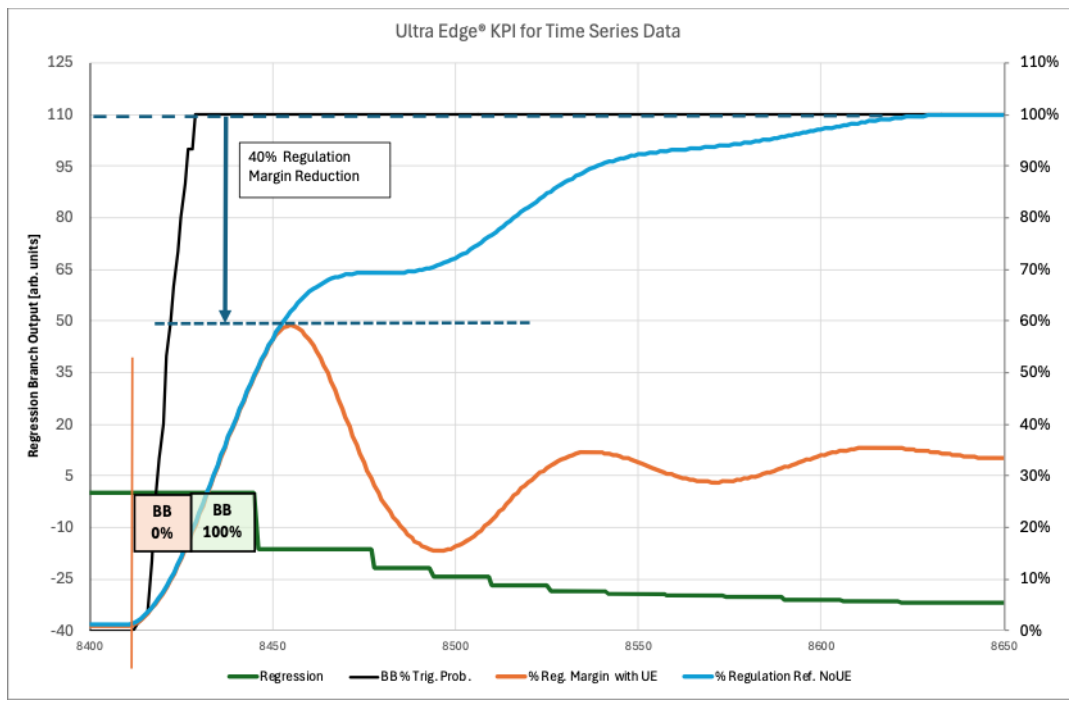

**Figure 5a.** Binary branch (800us) triggered with 99% binary branch accuracy.

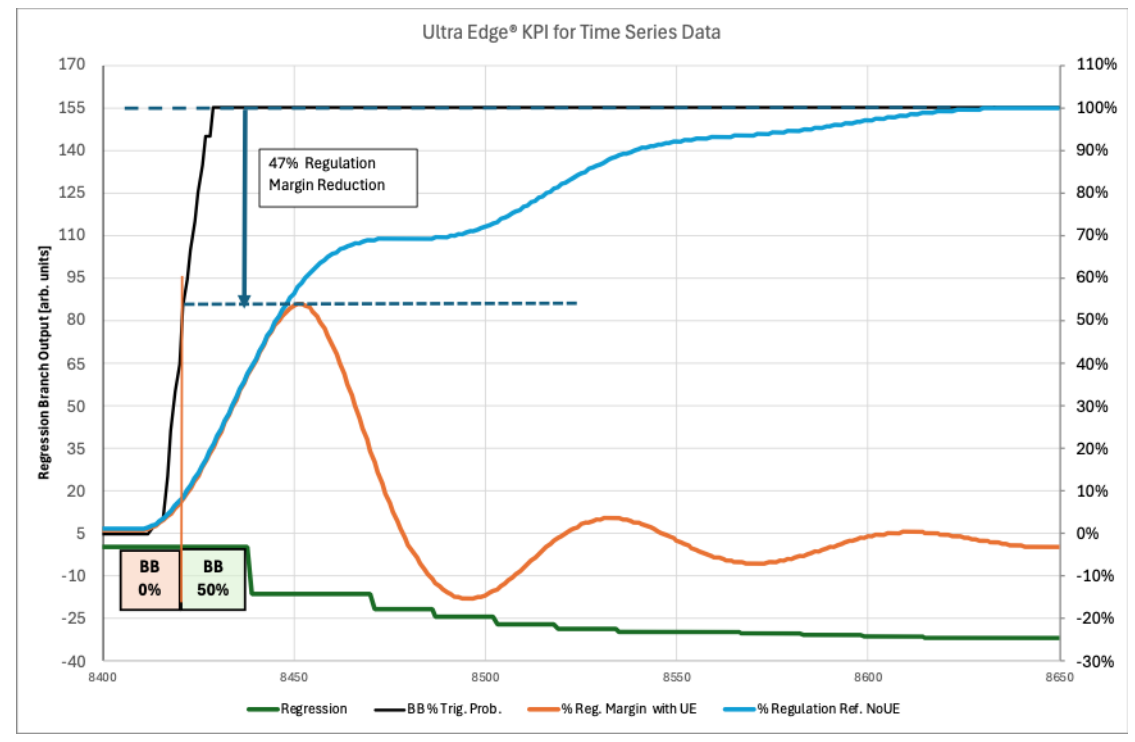

**Figure 5b.** Binary branch (800usec) triggered with 50% binary branch accuracy

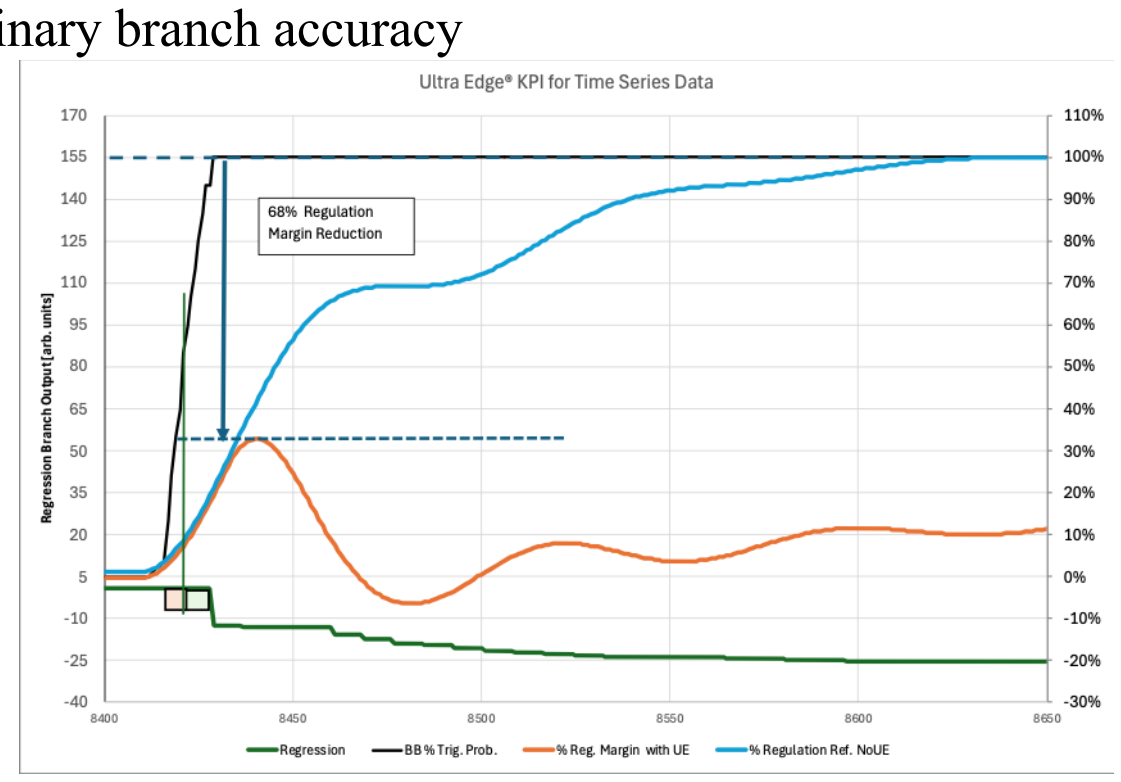

**Figure 5c.** Binary branch (200usec) triggered with 50% binary branch accuracy

The system is designed to be very flexible by leveraging Volumio functions for WAV playback. The WAV files generated via python for this project contain speed reference in the left channel and torque demand in the right channel. This configuration ensures synchronized control of both speed and torque, allowing the use of WLTP-compliant load profiles. We have also implemented a simple direct torque control device with 3-bit digital signal that is connected to power devices which is capable of generating 7 distinct torque steps. The results presented in this paper are limited to this simple direct torque control (trq1 – trq7).

Motor control datasets are challenging to analyze due to the nonlinear dynamics of electric machines, strong correlation between control variables, dynamic transient operating conditions, measurement noise, and temporal dependencies. One of the biggest barriers with integrating machine learning algorithms in motor control is the requirement for large amounts of well annotated data, whether it be used for training or post deployment data analysis. For a more optimal ML integrated motor control solution, we have shifted the emphasis from data volume to data effectiveness. In the quest to extract controller representative patterns from smaller sets of data, we leveraged Principal Component Analysis (PCA) and reduce the dimensionality of data. This work builds upon a widely adopted data analysis algorithm while proposing a novel way of correlating informative patterns to gain deeper insight into dynamical system response.

In the following section, we use PCA as a real-time motor

weeteq

control Key Performance Indicator (KPI). PCA is applied to quantify and visualize how Ultra Edge mitigates system nonlinearities in real-time within the theoretical framework of Lyapunov stability.

## III. Results

The lab setup is used for real-time demonstration using a python based GUI logging the data continuously as shown in Figure 4. The continuous data logging is only used for demonstration purposes. The data flow previously discussed, does not require any external software and solely relies on the data in the circular buffers and processing capabilities of the on-chip cores available within the weeteq Motor Controller IC. The GUI is useful for data visualization and the screen capture shows trq1-tr7 transition followed by trq7-tr1 transition. This torque transition is repeated with the first set displaying data with Ultra Edge disabled and the second set displaying data with Ultra Edge enabled. The speed error during the torque transients is reduced when Ultra Edge is enabled without changing the configuration of the main FOC controller. The data is shown at the bottom (labeled Display Manager – 2) with normalized post-processing block output acting on the outer speed control loop.

We have used the data from circular buffers of the Motor Controller IC for the remaining results discussed in this section. Since the circular buffers only capture data around the significant events, the analysis of data performed on chip is re-created in Figures 5a, 5b, and 5c, as this function is carried in real-time on-chip without continuous data logging. The time series data is not used as the basis for analysis except for the discussions around time synchronization of real-world events and model inference delays.

The model inference performance in the literature is typically discussed in terms of model accuracy. In real-time motor control applications, this is less critical for the system response correction compared to the latency of response. The impact of time delays in the system response is illustrated in Figure 5a for a motor control example with 800usec model inference period, 99% model accuracy, and 500usec main controller response time. In this example, the x-axis represents the data pointer updated at controller sample rate. Two different time synchronization scenarios are considered in terms of binary branch trigger probability with respect to the inference start time. The square boxes illustrate model inference period where the first scenario aligns with greater than 99% probability and the second scenario aligns with 50% probability which provides trigger of the significant event 320usec earlier. To reduce the overall regulation margin, the model inference time period must be less than 200usec as illustrated in Figure 5c.

Time series data work well for data visualization, however, a better tool for on-chip, unsupervised model performance indicator is the use of principal component analysis for evaluating the model performance during the transient response. This allows for reduction of multi-dimensional vector space and identification of steady state conditions and the optimum trajectory between two steady state conditions. In the PCA plot, departure from a steady state can easily be analyzed with a k-means cluster method. The steady state conditions appear as clusters as shown in Figure 6.

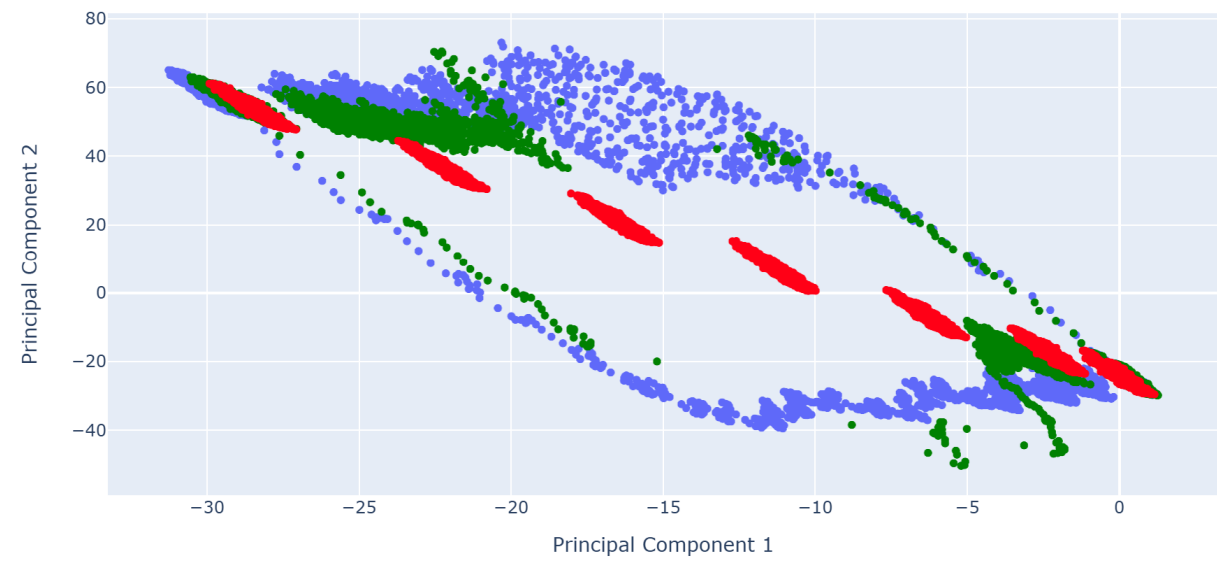

**Figure 6.** Principal Component Analysis of steady state and transients

The red clusters in the figure correspond to data taken during steady state operation with the constant torque setting (trq1, trq2, …trq7). Transient data for trq1 to trq7 transitions (data above the red clusters) and trq7 to trq1 transitions (data below the red clusters) are shown in blue and green. The data in Figure 6 was taken with mechanically coupled JK42BLS04-X028ED JKONGMOTOR BLDC motors for lightweight demonstration.

Plot of Figure 6 represents the first two principal components of PCA transform which are plotted on the X and Y axes. Through this visualization, it can be observed that the principal components follow a bounded trajectory from one steady state cluster to another steady state cluster. This trajectory is a representative of the system response in terms of principal components due to external events such as torque transient.

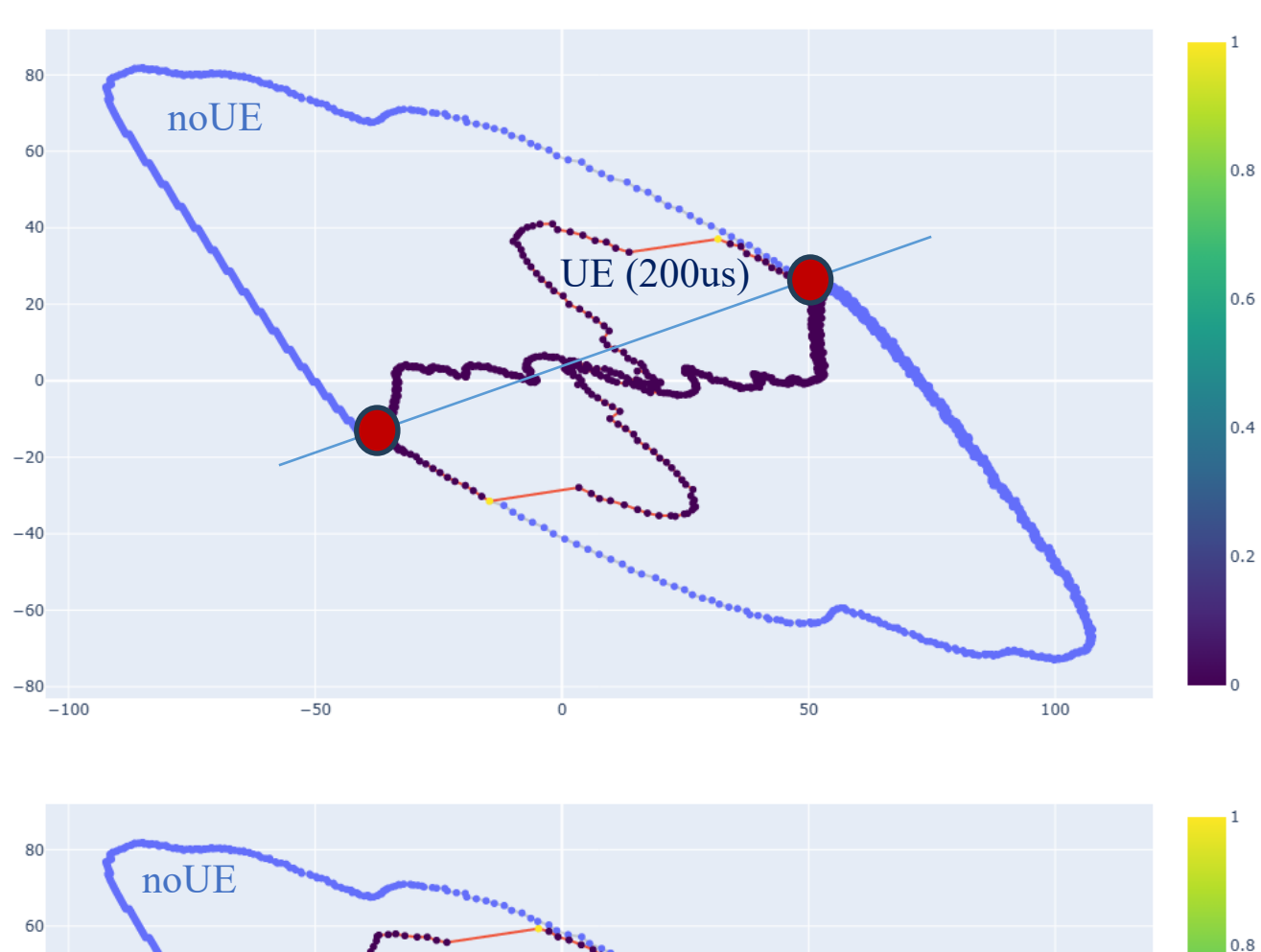

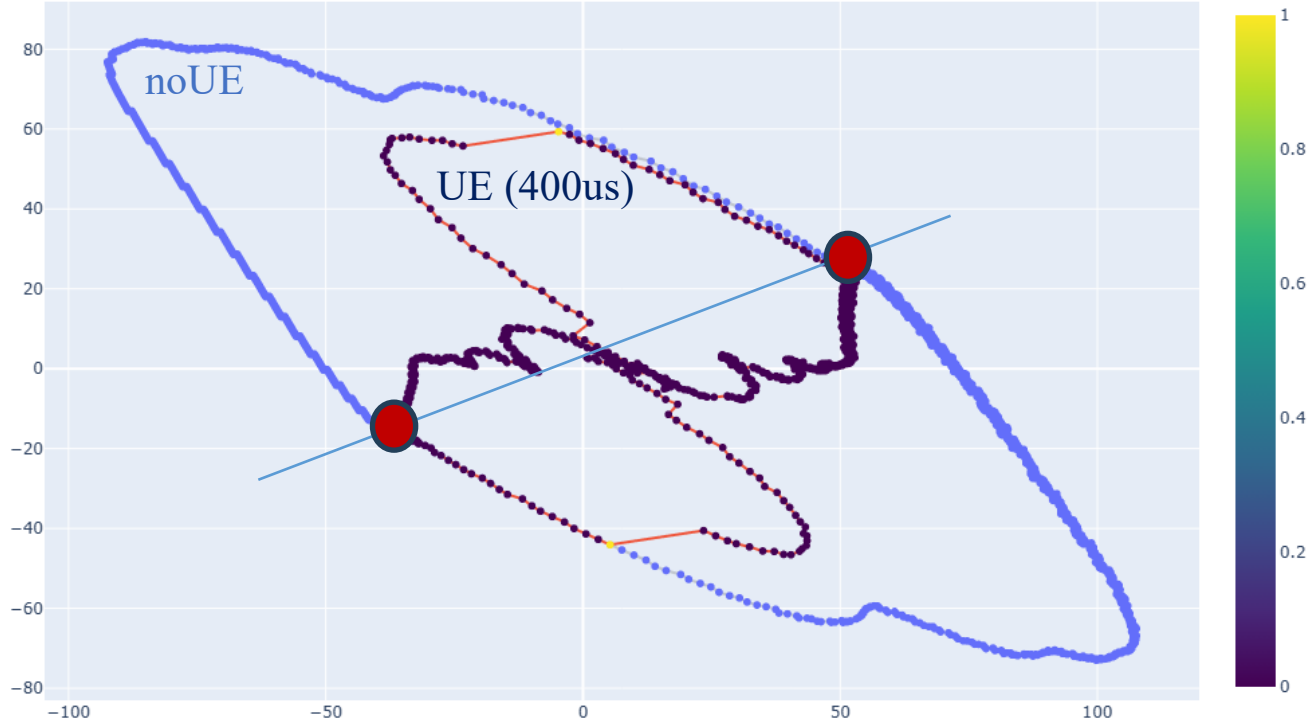

**Figure 7.** Simulated complete torque transients in both directions with 200usec trigger (top graph) and 400usec trigger (bottom graph). The straight line represents ideal transient path (KPI=0) between stead state conditions shown with red circles

weeteq

As indicated, the red clusters in Figure 6 are collected as various steady state conditions with stepwise incremental torque data, after sufficient time allowed between the torque steps. In Figure 7 the corresponding steady state data is marked with red circle. The ideal transient between these two steady state operating conditions follow a straight line and can be taken as the linear response that remain within the gain/bandwidth constraints of the control loop.

The benefits of PCA analysis become clearer when we consider the unsupervised analysis of model performance and use of trajectory path as key performance indicators (KPI). We defined the figure of merit as the area between the transient trajectory path and the ideal transient path where the ideal transient path is the straight line between two steady state clusters. The KPI of transient system response correction for various solutions is calculated as the ratio of the area of the corrected transient trajectory to the area of the uncorrected transient trajectory. While the KPI defined in this manner provides relative merit for each use case, it provides quantitative criterion to determine if any real-time system response correction is needed. Ideal performance would have a KPI value of 0.0, which is representative of the slowly changing steady state conditions resulting from a slowly varying torque change, that stays strictly within the linear response and gain/bandwidth constraints. It can be intuitively stated that the Ultra Edge response trajectory path, that is contained between the transient trajectory of a stable main controller response (without Ultra Edge correction) and the straight line representing the infinitesimal torque transient steps provides a stable response.

In this paper, the discussion is limited to the observations of the correlation between time series data and the trajectory path of principal components of the PCA transform. The corresponding KPI from PCA and the reduction of regulation margin from time series data is provided in Table 1 for the baseline model and Ultra Edge model with different model inference times. Baseline regulation margin is typically less than 5% and NRMR of 60% indicates better than 2% dynamic regulation margin for the overall system.

TABLE I

KPI AND REGULATION MARGIN COMPARISON

| trq1 to trq7 transient | KPI From PCA | Normalized Regulation Margin Reduction (NRMR) |
|---|---|---|
| Baseline (noUE) | 1.00 | 100% |
| UE (400us) | 0.37 | 50% |
| UE (200us) | 0.25 | 64% |

| trq7 to trq1 transient | KPI from PCA | Normalized Regulation Margin Reduction (NRMR) |
|---|---|---|
| Baseline (noUE) | 1.00 | 100% [1] |
| UE (400us) | 0.43 | 43% |
| UE (200us) | 0.20 | 67% |

The experimental data from weeteq Motor Control IC (Clyde IC) is limited to the minimum requirements around the significant events and do not show the complete transition between the two steady state clusters. To illustrate the complete transient trajectory path for the torque transients between steady state conditions, the data for the torque transients in Table 1 is plotted in Figure 7 using simulated data of the Clyde IC digital design comparing Ultra Edge® and Baseline (noUE) transients.

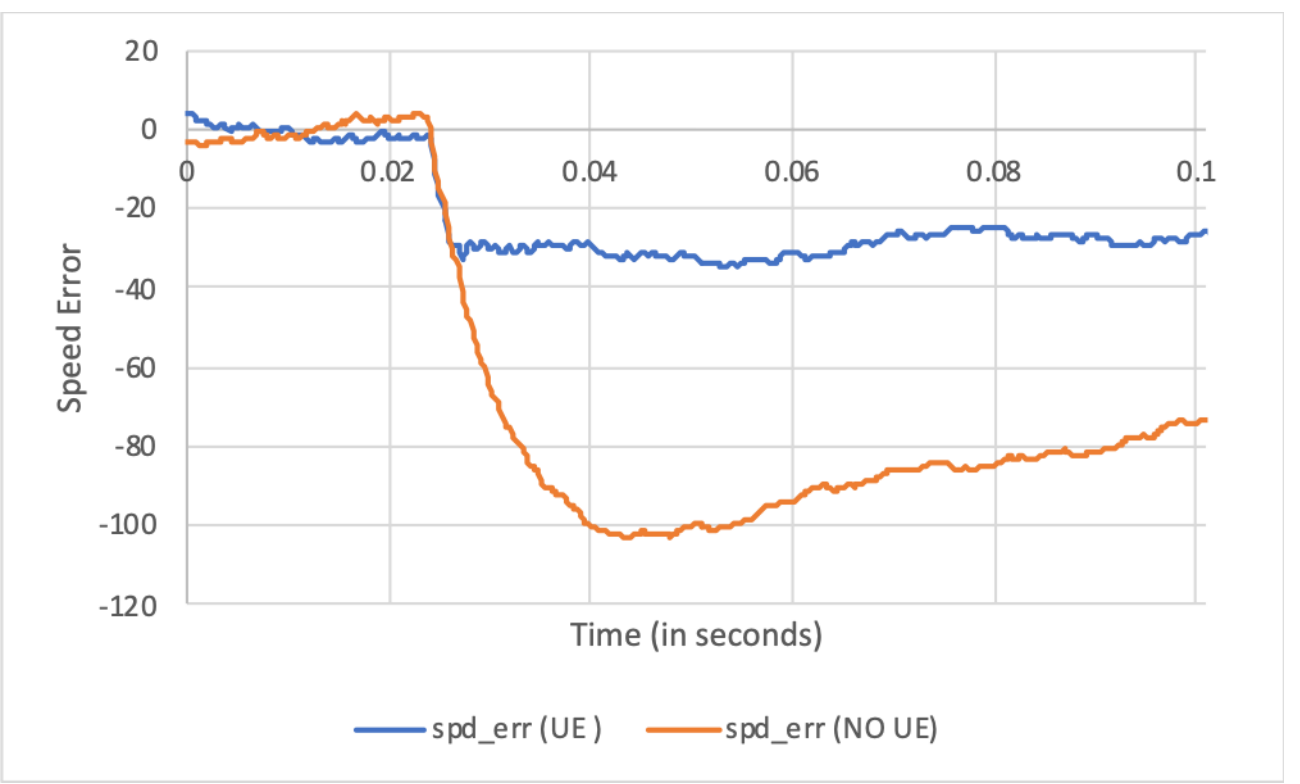

**Figure 8.** Time series data from Clyde IC with and without Ultra Edge correction

Figure 7, illustrates the system response correction of the baseline controller response (noUE: Ultra Edge disabled) to the trajectories of the models with different model inference times.

The on-chip transient response correction follows the same approach but the data from the circular buffers capture the start of the significant event and first part of the system response correction. This is plotted in Figure 8 and Figure 9 for time series data and the corresponding PCA plot respectively. The time series data shows 65% reduction in dynamic regulation error. In this example, the model inference period is 200us based on an input data sampling time of 50us. The data stored in the circular buffer for analysis is sampled at 200us intervals. Using similar considerations as illustrated in Figure 5c, we would expect to see Ultra Edge response correction within 400usec with some additional delays, including the controller response and measurement delays. The regression branch performance can be observed by the transient trajectory of the data points following the trigger of the significant event detection in Figure 9. The transient trajectory is completed within 200usec and the system response is already in steady state where the baseline main controller model can take over the system response operating close to the target steady state operation.

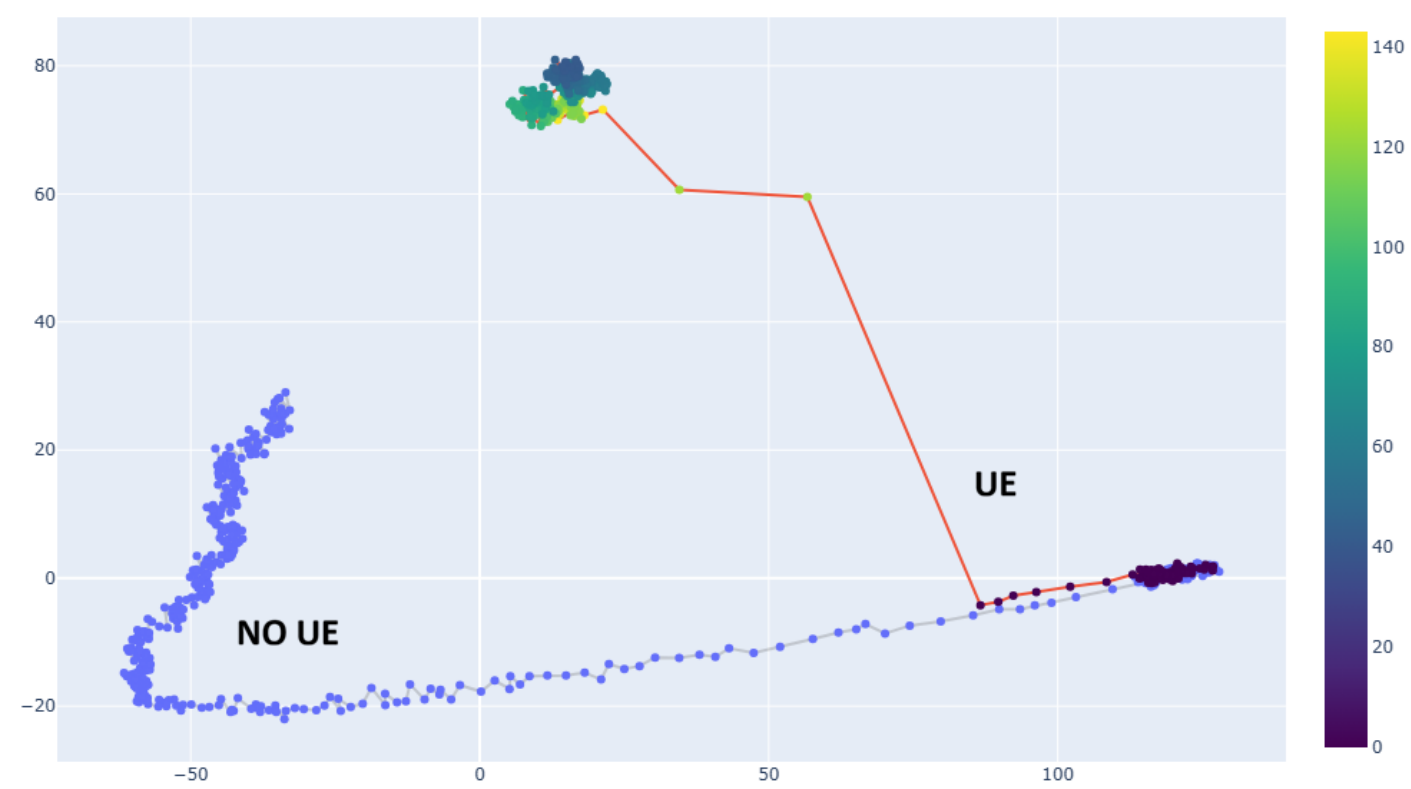

**Figure 9.** PCA of Clyde IC on-chip data with and without Ultra Edge correction

## IV. Conclusion

Ultra Edge® supplementary controller for real-time on-chip transient system response correction is developed and implemented in a motor controller integrated circuit. Methodology is developed for evaluating a figure of merit using only on-chip data captured around a significant event that is triggered by a lightweight binary branch. Ultra Edge® solution can be integrated within microcontrollers, and it is compatible with the real-time demands of motor controller systems. It is especially suitable for large signal transient response correction without disrupting the operation of the linear controllers during steady state operation. While we presented the results of Ultra Edge® for motor controller applications, we were able to apply Ultra Edge® technology to control systems of power inverters both in dq reference frame (stationary reference signals) and in abc reference frame (three phase sinusoidal voltage and current signals).

The data generated by the binary branch and regression branch during significant events provide an accurate real-time representation of large signal controller behavior that also correlates with the health of the equipment and external factors. This data is provided as Dynamic Digital Shadow which can be used to recreate these significant events for simulations and system level optimization offline. An interesting use case is the ability to maintain operation of the controller within the target specification, identifying the required configuration modifications, and reporting this significant event which would not be detected by IoT sensors connected to the system. In this scenario, Ultra Edge® corrects the system response in real-time and reports what would have happened without this correction. The ability of Ultra Edge® to reconstruct ideal trajectories using PCA inverse transforms with limited real-time data captured around significant events facilitates on-chip training and continuous monitoring of the real-time data at the control loop sample rates (for both outer loops and inner current loops of field oriented control solutions). Compared to IoT sensor-based health monitoring systems, amount of data required to capture the significant events is reduced by 10,000 times with 1ms-10ms resolution and 100% coverage of the real-time data sampled at control loop data rates. Deployment of these solutions in the field will enable an entirely new set of data, currently inaccessible at these rates, to the system level optimization platforms..

## Acknowledgment

Access the ARM IP was made possible by Arm Flexible Access Program for Startups. Motor Con-troller IC was developed using Synopsys EDA Tools and Synopsys Cloud Platform, Mathworks Simulink, Siemens Calibre tools and manufactured at TSMC with in-kind contributions of these Silicon Catalyst technology partners.